\documentclass[conference]{IEEEtran}
\IEEEoverridecommandlockouts
\usepackage{cite}
\usepackage{amsmath,amssymb,amsfonts}
\usepackage{algorithmic}
\usepackage{graphicx}
\usepackage{textcomp}
\usepackage{xcolor}
\usepackage{hyperref}  
\usepackage{wrapfig}
\def\BibTeX{{\rm B\kern-.05em{\sc i\kern-.025em b}\kern-.08em
    T\kern-.1667em\lower.7ex\hbox{E}\kern-.125emX}}
\begin{document}

\title{ANN-Based Grid Impedance Estimation for Adaptive Gain Scheduling in VSG Under Dynamic Grid Conditions}
\author{
    \IEEEauthorblockN{Quang-Manh Hoang,
    {Van Nam Nguyen}, 
    Taehyung Kim,
    Guilherme Vieira Hollweg,
    Wencong Su, 
    Van-Hai Bui{*}}
    
    \IEEEauthorblockA{Department of Electrical and Computer Engineering, University of Michigan-Dearborn, USA}
    
}
\maketitle

\begin{abstract}
In contrast to grid-following inverters, Virtual Synchronous Generators (VSGs) perform well under weak grid conditions but may become unstable when the grid is strong. Grid strength depends on grid impedance, which unfortunately varies over time. In this paper, we propose a novel adaptive gain-scheduling control scheme for VSGs. First, an Artificial Neural Network (ANN) estimates the fundamental-frequency grid impedance; then these estimates are fed into an adaptive gain-scheduling function to recalculate controller parameters under varying grid conditions. The proposed method is validated in Simulink and compared with a conventional VSG employing fixed controller gains. The results demonstrate that settling times and overshoot percentages remain consistent across different grid conditions. Additionally, previously unseen grid impedance values are estimated with high accuracy and minimal time delay, making the approach well suited for real-time gain-scheduling control.
\end{abstract}


\section{Introduction}
The increasing penetration of renewable energy sources (RES) has led to significant changes in power-system dynamics, necessitating advanced control strategies for inverter-based resources (IBRs) \cite{hoang2024decoupled, tian2024full}. Unlike conventional synchronous generators, IBRs rely on power electronics to interface with the grid, which alters overall system stability characteristics. Two common control paradigms for IBRs are grid-following (GFL) inverters and grid-forming (GFM) inverters. While GFL inverters operate by tracking the grid voltage and injecting the requisite active and reactive power, they struggle under weak-grid conditions due to their dependence on external synchronization signals. In contrast, GFM inverters, particularly those based on the Virtual Synchronous Generator (VSG) concept which emulate the dynamic behavior of traditional synchronous machines, thereby enhancing system stability and providing inertia support. However, despite their advantages in weak-grid scenarios, VSGs can become unstable under strong-grid conditions if their controllers are not properly designed.

Several studies have focused on the design of VSG power controllers \cite{Liu7182342, li2023improved, wu2016small, zhong2010synchronverters}. In \cite{Liu7182342}, the transient responses of VSG and droop-based GFM were compared, showing that the VSG’s larger effective inertia yields superior frequency stability. However, both the inertia and damping ratio were fixed for a single, constant grid impedance. In \cite{li2023improved}, the authors proposed an active-power control scheme to enhance VSG transient stability; nonetheless, they treated grid impedance as a static parameter, which limits applicability under realistic, time-varying conditions. Reference \cite{wu2016small} introduced a systematic four-step procedure for designing VSG power-loop parameters based on a line-frequency–averaged small-signal model that quantifies when active and reactive power loops may be decoupled. Although this analysis addresses active–reactive coupling, it assumes a fixed, predominantly inductive grid impedance and does not account for variations in the X/R ratio and their impact on loop coupling.

To overcome those drawbacks, many papers have concentrated on online estimation of grid impedance to improve control efficiency \cite{mohammed2022online, hoffmann2013minimal, qiu2022artificial, alves2018real, mohammed2022adaptive, suarez2022grid, cheng2023zero}. These methods are classified into two main categories: invasive and non-invasive methods. Invasive methods deliberately inject a known test signal into the network and measure the voltage or current response to compute impedance \cite{cheng2023zero, mohammed2022adaptive, suarez2022grid, mohammed2022online}. These algorithms have fast convergence; however, injected signals may disturb normal operation or violate grid codes. They may also require a dedicated injection interface. Specifically, in \cite{mohammed2022online}, the authors proposed an adaptive control strategy based on an online grid-impedance estimation algorithm. However, this estimation method has a large time delay. Moreover, such an intrusive method of injecting a 75 Hz signal into the VSG control loop may introduce unwanted interactions with the control dynamics. In \cite{hoffmann2013minimal}, the Extended Kalman Filter is used to estimate grid impedance in inductive–resistive power networks. However, the performance of this method depends on the initial condition guesses and the tuning of the process-noise covariance. Therefore, the authors in \cite{qiu2022artificial} proposed an ANN method to identify grid impedance with high accuracy and minimal time delay. However, this method was applied only to GFL inverters and did not consider a comprehensive range of grid conditions.

To overcome these disadvantages, this paper proposes an adaptive gain-scheduling control scheme which first determines the grid impedance online using an ANN model. After that, the adaptive gain-scheduling function is utilized to set the VSG parameters based on grid impedance information corresponding to weak, strong, and very strong grid conditions. The proposed method is simulated in MATLAB/Simulink, and the results demonstrate that the oscillations of active and reactive power are reduced compared to the case when the VSG operates with fixed gain values. Furthermore, control performance indices such as settling time and percentage overshoot are remained the same under various grid conditions. In addition, the ANN-based grid impedance estimator performs well under different, previously unseen grid conditions, with high accuracy and minimal delay.

The remainder of the paper is organized as follows. Section II presents the small-signal stability analysis of the VSG. Section III introduces the proposed adaptive gain-scheduling control scheme for the VSG under different grid conditions. In Section IV, the method is validated through MATLAB/Simulink simulations under various grid conditions. Finally, conclusions are drawn in Section V.

\section{Small signal stability issues of VSG under strong grid condition}

\begin{figure}[!h]
\centerline{\includegraphics[scale = 1]{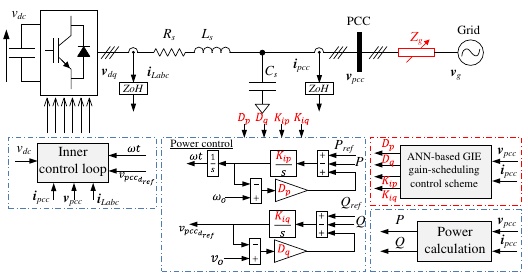}}
\caption{Configuration of the studied system.}
\label{fig:Config}
\end{figure}
There are several VSG topologies in the literature. In this paper, the VSG structure is similar to those proposed in \cite{zhong2010synchronverters, wu2016small} and is shown in Fig. \ref{fig:Config}.
The active-power reference consists of the setpoint, $P_{ref}$, plus a droop term. The droop power is calculated by multiplying the difference between the actual angular frequency of the capacitor voltages, $\omega$, and the nominal angular frequency, $\omega_0$, by a droop coefficient $D_p$. Since $ \omega \approx \omega_g$, the VSG can automatically change its output active power according to the grid frequency, $\omega_g$, thus implementing the $P -\omega$ droop mechanism. Meanwhile, the integral term $K_{ip}/s$ is adopted here to introduce the virtual inertia. The reactive-power controller operates on the same principle as the active-power controller.

 To identify the proper VSG gains, it is crucial first to analyze the small-signal stability of the VSG under varying grid conditions. The active- and reactive-power flow equations between the inverter and the grid can be expressed as: \\
\vspace{-0.1cm}
\begin{equation}\label{eq_Active}
    P_{pcc} = \frac{3}{R_g^2 + X_g^2} \left(R_g V_{pcc}^2 - R_g V_{pcc} V_g \cos \delta + X_g V_{pcc} V_g \sin \delta \right),
\end{equation}
\begin{equation}\label{eq_ReActive}
    Q_{pcc} = \frac{3}{R_g^2 + X_g^2} \left(X_g V_{pcc}^2 - X_g V_{pcc} V_g \cos \delta - R_g V_{pcc} V_g \sin \delta \right),
\end{equation}
where, $Z_g = R_g + jX_g$ represents the grid impedance, $\delta$ is the phase angle difference between $v_{pcc}$ and $v_g$. $V_{pcc}$ and $V_g$ denote the RMS voltages of point-of-common coupling (PCC) bus and grid, respectively. To analyze the VSG dynamics, the small-signal models of (\ref{eq_Active}) and (\ref{eq_ReActive}) are derived using the Jacobian linearization method from \cite{tian2024full}.
\begin{equation}\label{eq_Jacobian}
    \begin{bmatrix}
    \Delta P_{\mathrm{pcc}} \\[4pt]
    \Delta Q_{\mathrm{pcc}}
    \end{bmatrix}
    =
    \begin{bmatrix}
    A & B \\[4pt]
    C & D
    \end{bmatrix}
    \begin{bmatrix}
    \Delta \delta \\[4pt]
    \Delta V_{pcc}
    \end{bmatrix}
\end{equation}
where $A$, $B$, $C$, and $D$ are the partial derivatives of $P_{pcc}$ and $Q_{pcc}$ evaluated at the operating point. In this paper, the coupling terms $B$ and $C$ between active and reactive power are ignored when designing the controllers. Therefore, the relationships $P_{pcc} - \delta$ and $Q_{pcc} - \Delta V_{pcc}$ are presented as follows:
\begin{equation}\label{eq_SmallP}
    \frac{\Delta P_{pcc}}{\Delta \delta} = A = \frac{3}{R_g^2 + X_g^2} \left(R_g V_{pcc_0} V_g \sin \delta_0 + X_g V_{pcc_0} V_g \cos \delta_0 \right),
\end{equation}
\begin{equation}\label{eq_SmallQ}
    \frac{\Delta Q_{pcc}}{\Delta V_{pcc}} = D = \frac{3}{R_g^2 + X_g^2} \left(X_g V_{pcc_0} V_g \sin \delta_0 - R_g V_{pcc_0} V_g \cos \delta_0 \right),
\end{equation}
where the subscript $0$ denotes the steady-state operating point. The transfer function of active and reactive power control are shown as follow:
\begin{equation}\label{eq_ControllawP}
    \frac{\Delta \delta}{\Delta P_{ref}} = \frac{K_{ip}}{s(s + D_pK_{ip})},
\end{equation}
\begin{equation}\label{eq_ControllawQ}
    \frac{\Delta V_{pcc}}{\Delta Q_{ref}} = \frac{K_{iq}}{s + D_qK_{iq}},
\end{equation}
where $D_p$ and $D_q$ are the droop coefficients, and $K_{ip}$ and $K_{iq}$ determine the virtual inertia of the VSG. From (\ref{eq_SmallP})–(\ref{eq_ControllawP}) and (\ref{eq_SmallQ})–(\ref{eq_ControllawQ}), the open-loop transfer functions for the active and reactive power control loops are denoted as $G^P_{op}(s)$ and $G^Q_{op}(s)$, which are presented in (\ref{eq_OPloopP}) and (\ref{eq_OPloopQ}), respectively.
\begin{equation}\label{eq_OPloopP}
\begin{split}
  G^P_{op}(s)
  &=\frac{\Delta P_{pcc}}{\Delta P_{ref}} =  \frac{K_{ip}}{s(s + D_pK_{ip})}
     \;\cdot\;
     \frac{3}{R_g^2 + X_g^2} \\
  &\cdot
  \Bigl(
    R_g\,V_{pcc_0}\,V_g\,\sin\delta_0
    +\,X_g\,V_{pcc_0}\,V_g\,\cos\delta_0
  \Bigr)\,. 
\end{split}
\end{equation}
\begin{equation}\label{eq_OPloopQ}
\begin{split}
  G^Q_{op}(s)
  &=\frac{\Delta Q_{pcc}}{\Delta Q_{ref}} =  \frac{K_{iq}}{s + D_qK_{iq}}
     \;\cdot\;
     \frac{3}{R_g^2 + X_g^2} \\
  &\cdot
  \Bigl(
    X_g V_{pcc_0} V_g \sin \delta_0 - R_g V_{pcc_0} V_g \cos \delta_0
  \Bigr)\,. 
\end{split}
\end{equation}

\begin{figure}[!h]
\centerline{\includegraphics[scale = 0.8]{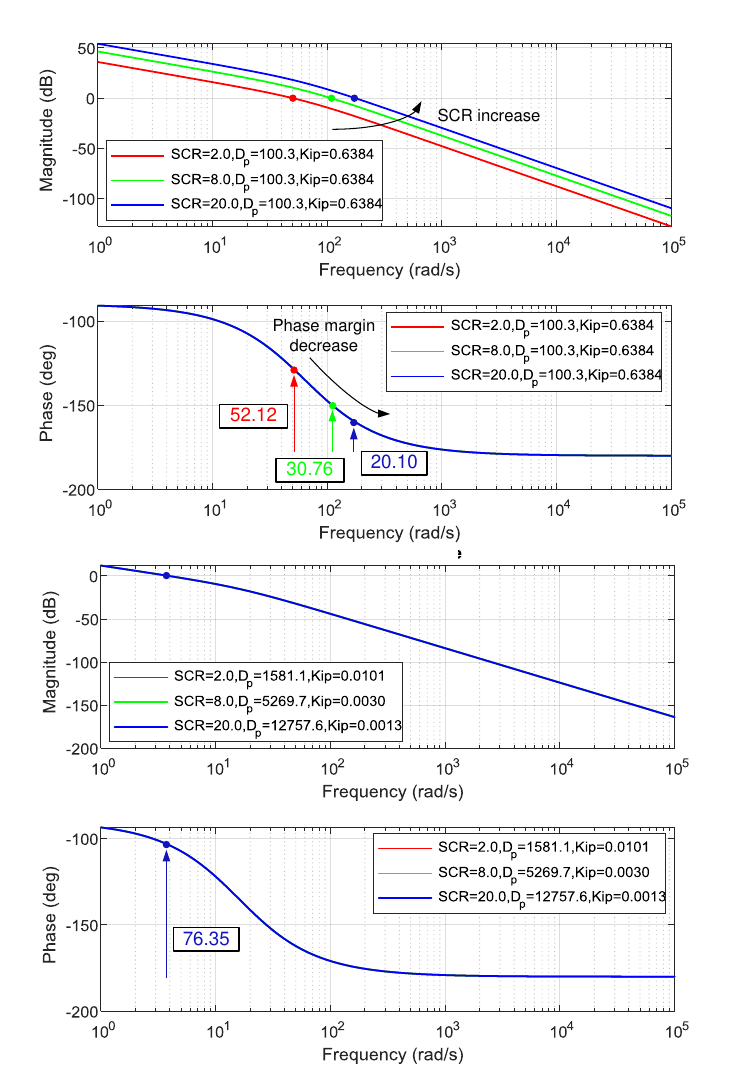}}
\caption{Bode diagram of $G^P_{op}(s)$ with fix (upper) and adjustable (lower) parameters.}
\label{fig:Bode}
\end{figure}

As shown in (\ref{eq_OPloopP}) and (\ref{eq_OPloopQ}), the dynamics of the VSG depend on $D_p$, $D_q$, $K_{ip}$, $K_{iq}$, $L_g$, and $R_g$. To examine the dynamics of the VSG, the Bode diagram is plotted under different scenarios, as shown in Fig.~\ref{fig:Bode}. The upper picture illustrates that, with fixed VSG parameters, a stronger grid (higher SCR) reduces the phase margin, potentially causing oscillations or instability \cite{tian2024full}. However, by adjusting the VSG parameters based on grid impedance, the phase margin can be maintained across various scenarios.

Determining how and when to adjust VSG parameters remains a critical challenge, particularly under varying grid conditions. To address this, an adaptive gain-scheduling control scheme is proposed, which dynamically adjusts the VSG parameters based on real-time grid impedance estimation.
\section{Proposed Adaptive Gain-Scheduling Control Scheme for VSG}

\begin{figure*}[htbp]
  \centering
  \includegraphics[width=0.9\textwidth]{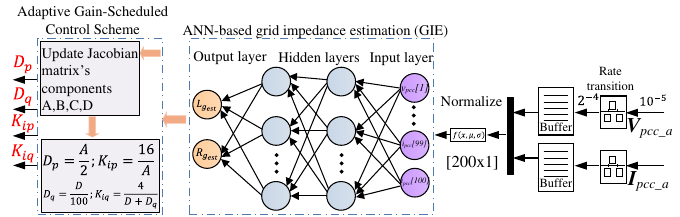}  % thay bằng tên file ảnh của bạn
  \caption{Proposed control architecture.}
  \label{fig:Adaptive}
\end{figure*}

Fig.~\ref{fig:Adaptive} illustrates the structure of the proposed method. Only the steady-state grid impedance is considered, and the transient period is relatively short; therefore, only $v_{pcc}$ and $i_{pcc}$ are used as inputs to the ANN model. If the grid impedance estimation performance during the transient period needs to be considered, the VSG parameters should also be included as inputs. After obtaining the fundamental-frequency grid impedance values $L_g$ and $R_g$, these are used in the adaptive gain-scheduling control block to adjust the VSG gains, ensuring control performance under dynamic conditions.

\begin{table}
\caption{ANN model parameters}\label{tb:ANN}
\centering
    \begin{tabular}{|c|c|}
                \hline \hline
                Parameters & Values \\ \hline
                Algorithm & \shortstack{Levenberg- \\ Marquardt} \\ \hline
                Max epochs & 500 \\ \hline
                Goal (MSE) & $10^{-5}$ \\ \hline
                Activation fcn & tansig \\ \hline
                Damping factor $\mu$ & $10^{-6}$ \\ \hline
                Hidden neurons & $8$ \\ \hline
                \hline
    \end{tabular}
\end{table}
\subsection{ANN-based Online Grid Impedance Estimation}
\begin{figure}[!h]
\centerline{\includegraphics[scale = 0.8]{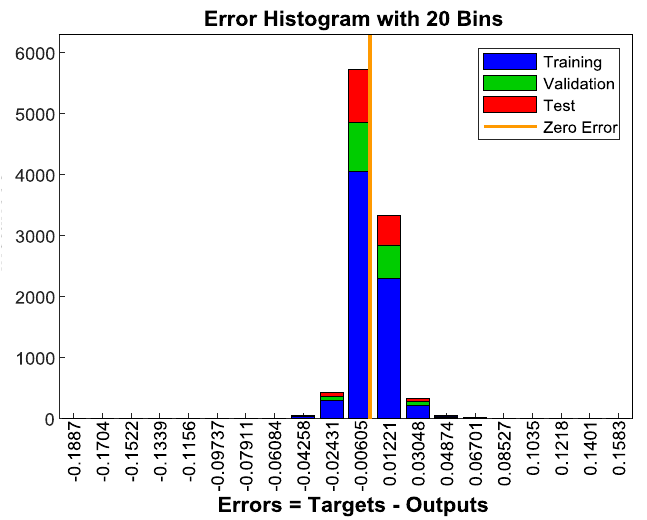}}
\caption{Neural Network Training Error Histogram.}
\label{ANN_error}
\end{figure}

\begin{figure}[!h]
\centerline{\includegraphics[scale = 0.8]{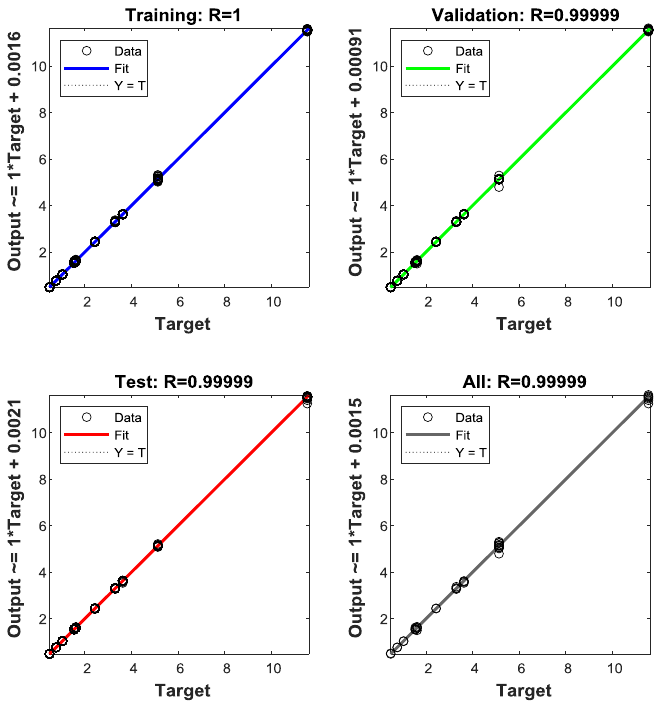}}
\caption{Neural Network Training Regression.}
\label{ANN_regression}
\end{figure}

\begin{figure}[!h]
\centerline{\includegraphics[scale = 0.8]{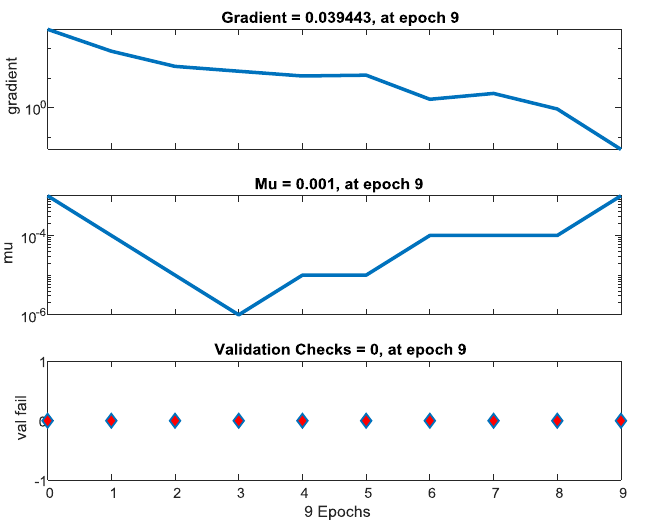}}
\caption{Levenberg-Marquardt Optimization Progress.}
\label{ANN_LM}
\end{figure}

The proposed ANN-based online grid-impedance estimation consists of two stages: data collection and ANN model training, validation, and testing. In the first stage, the system is simulated in MATLAB/Simulink under various operating conditions and grid impedance values (SCR = [2, 4.5, 7, 9.5, 15]). Next, a total of 5,000 samples are collected and split into training (70\%), validation (15\%), and testing (15\%) datasets. The ANN training parameters are summarized in Table~\ref{tb:ANN}. The training and validation results are shown in Fig.~\ref{ANN_error}, Fig.~\ref{ANN_regression}, and Fig.~\ref{ANN_LM}. The error histogram (Fig.~\ref{ANN_error}) displays the distribution of prediction errors with separate bars for the training, validation, and test sets. Most errors are distributed around zero, indicating that the network’s predictions are highly accurate for the vast majority of samples. The overlap of the validation and test histograms with the training histogram demonstrates good generalization and a lack of overfitting. Additionally, the approximate symmetry about zero confirms minimal systematic bias in the model’s estimates.

The regression plots in Fig.~\ref{ANN_regression} illustrate the linear relationship between the network outputs and the true target values for the training, validation, and test sets. In each case, the points lie tightly along the ideal line ($y = x$), with correlation coefficients ($R$) approaching 1 and negligible intercepts. This alignment confirms that the model achieves near-perfect predictive accuracy across all data splits and exhibits minimal bias. Training was performed using the Levenberg–Marquardt algorithm, which combines the rapid convergence properties of the Gauss–Newton method with the stability of steepest-descent \cite{matlab_trainlm}. For moderately sized feedforward networks, Levenberg–Marquardt typically requires a small number of epochs to reach a low training error, ensuring both efficiency and robust generalization. Specifically, the top picture in Fig.~\ref{ANN_LM} shows the gradient during the Levenberg–Marquardt optimization process. A decreasing gradient indicates that the algorithm is approaching a critical point of the loss function. The gradient approaches zero after only nine epochs, indicating that the network has converged or is near convergence in a very short time. The middle plot displays the values of $\mu$ during training, which balances the steepest-descent and Gauss–Newton update steps. $\mu$ remains within a reasonable range, never spiking excessively, indicating a well-behaved, stable training process. The bottom plot in Fig.~\ref{ANN_LM} shows the validation-check values, which increment each time the validation loss fails to improve. Here, the validation check remains at zero for all nine epochs, meaning the validation error never increased, indicating that the network either improved or maintained its generalization performance throughout. After obtaining the ANN model, it is utilized to predict new values of grid impedance in the Simulink environment.


Fig.~\ref{fig:Adaptive}, as previous discussed, illustrates the implementation of the proposed method in the Simulink environment. Due to a mismatch between the simulation sampling time ($10\ \mu\mathrm{s}$) and the ANN input sampling rate ($200\ \mu\mathrm{s}$), a rate transition block is used to resolve this issue. Since 200 points per cycle (0.02 s) must be collected before being fed to the ANN model, a buffer block accumulates input data for each estimation, introducing a one-cycle delay. After buffering, the data are normalized using the z-score method with the function:
\[
f(\mathbf{X},\mu,\sigma) = \frac{\mathbf{X} - \mu}{\sigma},
\]
where $\mu$ and $\sigma$ are the mean and standard deviation from the training phase. The ANN input vector is  
\[
\mathbf{X} = [v_{pcc}[1], \ldots, v_{pcc}[100],\, i_{pcc}[1], \ldots, i_{pcc}[100]]^T.
\]
After obtaining the grid impedance at the fundamental frequency, the adaptive gain-scheduling control function-introduced in the next section-needs to be implemented.

\subsection{Adaptive Gain-Scheduled Control Function}
Based on the estimated grid impedance, the Jacobian matrix components in (\ref{eq_Jacobian}) are first recalculated. The VSG parameters are then updated as previously shown and expressed as follows:

\begin{equation}\label{eq_PramsUpdateP}
    D_p = \frac{A}{2}; K_{ip} = \frac{16}{A},
\end{equation}
\begin{equation}\label{eq_PramsUpdateQ}
    D_p = \frac{D}{100}; K_{iq} = \frac{4}{D+D_q}.
\end{equation}

Detailed calculations are presented in the Appendix section. The proposed adaptive gain-scheduling scheme aims to maintain control performance under different grid conditions, as discussed in Section II, thereby improving VSG stability. The desired control performance targets are a settling time of $T_s = 1\ \mathrm{s}$ and zero overshoot. However, due to the omission of active–reactive coupling terms, there may be some mismatch between the design and validation phases. This discrepancy opens research questions for future work. Further analysis is provided in the next section.

\section{Simulation results}
To validate proposed method, model is simulated on MATLAB/Simulink with the parameters listed on the Table \ref{tab_SysPara}.
\subsection{Simulation scenarios}
        \begin{table}
		\caption{Parameters of the system}\label{tab_SysPara}
		\centering
		\begin{tabular}{p{1.5in} p{1in}}
			\hline\hline
			\textbf{Model parameters} \\ \hline
                Rated power & $S_{rated} = 5kVA$\\
			Sampling period   & $T_{sam}=50\mu{}s$\\
			dc bus voltage  & $u_{dc}=800V$\\
			LC-filter  & $L_s=1mH$\\
			&  $C_s=50\mu{}F$\\
			Switching frequency  & $f_s=10kHz$\\
			RMS grid voltage & $V_{g}=110V$\\
                \hline
			\textbf{Control parameters}  & \\ \hline
			Current control 	& \\
                Response time & $T_{cres}=1ms$\\
                Controller gains & $K_{pc} = 12.5664$\\
                    & $K_{ic} = 3.9478.10^4$\\
			Voltage control 	& \\ 
                Response time & $T_{vres}=10ms$\\
                Controller gains & $K_{pv} = 0.0628$\\
                    & $K_{iv} = 19.7392$\\
                Power control& \\
                Response time & $T_s = 1s$\\
                Controller gains & $D_p = 2.087. 10^3$\\
                    & $K_{ip} = 0.00767$\\
                    & $D_q = 0.687$ \\
                    & $K_{iq} = 0.115$\\
             \hline
		\end{tabular}
	\end{table}
The proposed method is validated through a 60 s MATLAB/Simulink simulation under three scenarios:  \\
- First 20 s: $SCR = 2$ (weak grid);  \\
- Middle 20 s: $SCR = 8$ (strong grid);  \\
- Final 20 s: $SCR = 20$ (stiff grid).  

At the midpoint of each scenario, the reference power is changed to examine the VSG’s transient response under varying grid conditions:  \\
- At 10 s: $P_{\text{ref}}: 2\,\text{kW} \rightarrow 2.5\,\text{kW}$; \\ 
- At 30 s: $P_{\text{ref}}: 2.5\,\text{kW} \rightarrow 3\,\text{kW}$;\\  
- At 50 s: $Q_{\text{ref}}: 1\,\text{kVAr} \rightarrow 1.5\,\text{kVAr}$.  

Furthermore, the proposed adaptive method (denoted $A\text{VSG}$) is compared with the conventional method (denoted $C\text{VSG}$), which uses fixed parameters.

\subsection{Simulation results and discussions}

\begin{figure*}[!h]
\centerline{\includegraphics[scale = 1.2]{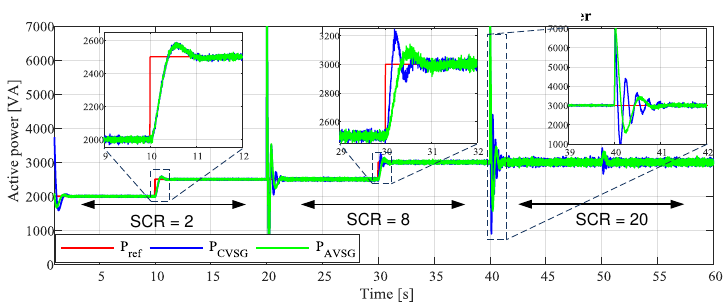}}
\caption{Active power of VSG with and without proposed adaptive control scheme.}
\label{fig:ActiveP}
\end{figure*}
\begin{figure*}[!h]
\centerline{\includegraphics[scale = 1.2]{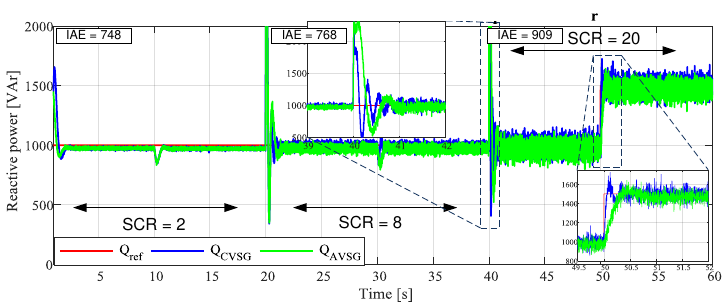}}
\caption{Reactive power of VSG with and without proposed adaptive control scheme.}
\label{fig:ActiveQ}
\end{figure*}

\begin{figure*}[!h]
\centerline{\includegraphics[scale = 1.2]{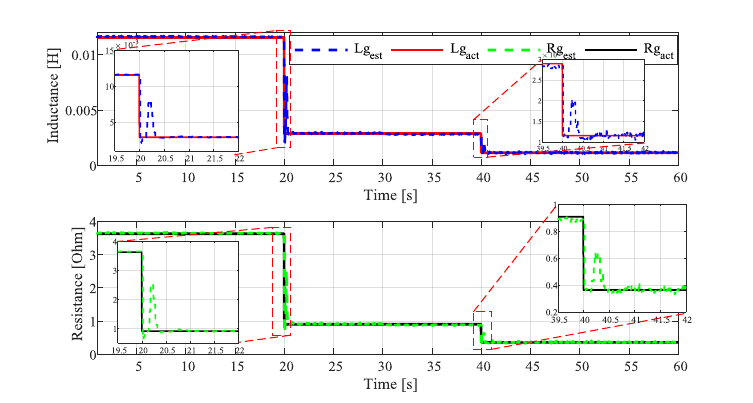}}
\caption{ANN-based online grid impedance estimation performance.}
\label{fig:GIE}
\end{figure*}

Simulation results, including active and reactive power and grid-impedance estimation, are shown in Fig.~\ref{fig:ActiveP}, Fig.~\ref{fig:ActiveQ}, and Fig.~\ref{fig:GIE}. As shown in Fig.~\ref{fig:ActiveP}, the VSG performs well for both $CVSG$ and $AVSG$ control schemes under a weak grid. When $P_{\mathrm{ref}}$ varies from $2\,\mathrm{kW}$ to $2.5\,\mathrm{kW}$, the transient responses of both schemes are similar, since they use the same values of $D_p$, $K_{ip}$, $D_q$, and $K_{iq}$ in this scenario. The settling time is $1\,\mathrm{s}$ as desired; however, the percentage overshoot is nonzero, due to ignoring the coupling terms of active and reactive power during the controller design phase. This will be addressed in future work.

As $SCR$ rises to 8, both the $CVSG$ and $AVSG$ methods respond to the change in active power setpoint, but power oscillations intensify at $t=30\,\mathrm{s}$ for the conventional method, as shown in Fig.~\ref{fig:ActiveP}. However, by adaptively recalculating VSG parameters in real time, the $AVSG$ control scheme achieves the desired settling time while maintaining the same percentage overshoot as in the previous period. Additionally, the chattering of the power signal increases due to the decrease in grid impedance, which acts as a low-pass filter component.

During the stiff-grid case ($SCR = 20$), severe oscillations precede stabilization, as seen in both Fig.~\ref{fig:ActiveP} and Fig.~\ref{fig:ActiveQ}. At $t = 40\,\mathrm{s}$, the SCR increases from 8 to 20, causing oscillations under the conventional control scheme. By using the proposed method, the oscillations are significantly reduced. At $t = 50\,\mathrm{s}$, the reactive power reference changes from $1\,\mathrm{kVAr}$ to $1.5\,\mathrm{kVAr}$; both control schemes reach the new setpoint. However, the conventional VSG does not guarantee the predefined control performance, whereas the adaptive VSG does.

The ANN-based estimator accurately tracks actual values with minimal delay (0.02\,s), significantly outperforming \cite{mohammed2022online} (Fig.~\ref{fig:GIE}). The zoom-in pictures show the tracking performance during both the steady state and the transient period of the ANN-based grid-impedance estimator. Under strong and stiff grid conditions, there are some peak errors in both $R_g$ and $L_g$ during the transient period. These errors appear because of the SCR values of [8, 20] represent grid conditions that the ANN had not encountered before. Moreover, because the inputs do not take into account the VSG parameters, the ANN does not fully capture grid impedance dynamics during the transient period.

\section{Conclusion}
This paper presented a novel ANN-based adaptive gain-scheduling control scheme for VSGs operating under varying grid conditions. Unlike conventional fixed-gain controllers, the proposed method dynamically adjusts VSG parameters based on real-time grid-impedance estimation. Key contributions include: first, a small-signal stability analysis of the VSG under different grid conditions; second, an ANN-based online grid-impedance estimation approach with high accuracy; and third, a gain-scheduled control scheme that guarantees VSG performance across weak and very strong grid conditions. Simulation results confirm that the proposed method effectively reduces settling time and oscillations while accurately estimating grid impedance.

\section*{Acknowledgment}
The authors work was supported by the University of Michigan-Dearborn’s Office of Research “Research Initiation and Development". The authors have made the Matlab code and simulation files associated with this paper publicly available at: \url{https://github.com/ManhqhUMich12/ANN_Based_GIE_AdaptiveControlGramework_VSG}.
\appendix                          % chuyển sang phần appendices
\section*{VSG parameters calculation}
\begin{itemize}
  \item Active power control loop:  
    The closed-loop transfer function of the active power control loop is presented as follows:
    \begin{equation}
      G^P_{cl}(s) = \frac{K_{ip}A}{s^2 + D_p K_{ip} s + K_{ip} A},
    \end{equation}
    which is a second-order transfer function with natural frequency $\omega_n = \sqrt{K_{ip}A}$, damping factor $\xi = \frac{D_p K_{ip}}{2\sqrt{K_{ip}A}}$, and settling time $T_s = \frac{4}{\xi \omega_n}$. Based on the desired characteristics (1 s settling time and no overshoot), let $\xi = 1$ and thus $\omega_n = 4$. The updated equations for $D_p$ and $K_{ip}$ are then
    \begin{equation}
      D_p = \frac{A}{2}, 
      \quad
      K_{ip} = \frac{16}{A}.
    \end{equation}

  \item Reactive power control loop:  
    The closed-loop transfer function of the reactive power control loop is presented as follows:
    \begin{equation}
      G^Q_{cl}(s) = \frac{K_{iq} D}{s + D_q K_{iq} + D K_{iq}}.
    \end{equation}
    Using the final value theorem, the steady-state value is 
    \[
      y_\infty = G^Q_{cl}(0) = \frac{D}{D_q + D}.
    \]
    For a unit step reference, the steady-state error is
    \begin{equation}
      e_\infty = 1 - y_\infty = \frac{D_q}{D_q + D}.
    \end{equation}
    To make the steady-state error negligible, choose 
    \[
      D_q = \frac{D}{100}.
    \]
    The settling time is $T_s = 4\tau$, where $\tau = \frac{1}{K_{iq}(D + D_q)}$. Setting $T_s = 1$ s gives
    \begin{equation}
      K_{iq} = \frac{4}{D + D_q}.
    \end{equation}
\end{itemize}

\bibliographystyle{IEEEtran}
\bibliography{IEEEabrv,bibliography}
%\bibliography{ECCEFullPaper}  % no “.bib” extension here

\end{document}